\documentclass[superscriptaddress,prb,amsmath,twocolumn,amssymb]{revtex4}

 \usepackage{natbib}
\usepackage{dcolumn}
\usepackage{amsfonts}
\usepackage{amssymb}
\usepackage{amsmath}
\usepackage{amsthm}
\usepackage{latexsym}
\usepackage{epsfig}
\usepackage{color}
\usepackage{lscape}
\usepackage{multirow}
\usepackage{braket}
\usepackage{bbm}
\usepackage{mathrsfs}
\usepackage{fancyhdr}
\usepackage{lastpage}
\usepackage{soul}
\usepackage{subfigure}
\usepackage[normalem]{ulem}
\usepackage[hidelinks=true]{hyperref}

\newcommand{\be}{\begin{equation}}
\newcommand{\ee}{\end{equation}}
\newcommand{\ba}{\begin{array}}
\newcommand{\ea}{\end{array}}
\newcommand{\bea}{\begin{eqnarray}}
\newcommand{\eea}{\end{eqnarray}}

\begin{document}

\title{9~GHz measurement of squeezed light by interfacing silicon photonics and integrated electronics}

\author{Joel~F.~Tasker}
\thanks{These authors contributed equally}
\affiliation{Quantum Engineering Technology Labs, H. H. Wills Physics Laboratory
and Department of Electrical \& Electronic Engineering, University of Bristol, BS8 1FD, United Kingdom}
\author{Jonathan~Frazer}
\thanks{These authors contributed equally}
\affiliation{Quantum Engineering Technology Labs, H. H. Wills Physics Laboratory
and Department of Electrical \& Electronic Engineering, University of Bristol, BS8 1FD, United Kingdom}
\affiliation{Quantum Engineering Centre for Doctoral Training, H. H. Wills Physics Laboratory and Department of Electrical \& Electronic Engineering, University of Bristol, BS8 1FD, United Kingdom}
\author{Giacomo~Ferranti} 
\affiliation{Quantum Engineering Technology Labs, H. H. Wills Physics Laboratory
and Department of Electrical \& Electronic Engineering, University of Bristol, BS8 1FD, United Kingdom}
\author{Euan~J.~Allen} 
\affiliation{Quantum Engineering Technology Labs, H. H. Wills Physics Laboratory
and Department of Electrical \& Electronic Engineering, University of Bristol, BS8 1FD, United Kingdom}
\author{L\'{e}andre~F.~Brunel}
\affiliation{Universit\'{e} C\^{o}te d`Azur, CNRS, Institut de Physique de Nice (INPHYNI), UMR 7010, Parc Valrose, Nice Cedex 2, France}
\author{S\'{e}bastien~Tanzilli}
\affiliation{Universit\'{e} C\^{o}te d`Azur, CNRS, Institut de Physique de Nice (INPHYNI), UMR 7010, Parc Valrose, Nice Cedex 2, France}
\author{Virginia~D'Auria}
\affiliation{Universit\'{e} C\^{o}te d`Azur, CNRS, Institut de Physique de Nice (INPHYNI), UMR 7010, Parc Valrose, Nice Cedex 2, France}
\author{Jonathan C. F. Matthews}
\email{Jonathan.Matthews@bristol.ac.uk}
\affiliation{Quantum Engineering Technology Labs, H. H. Wills Physics Laboratory
and Department of Electrical \& Electronic Engineering, University of Bristol, BS8 1FD, United Kingdom}

\date{\today}

\maketitle

\noindent\textbf{
Photonic quantum technology can be enhanced by monolithic fabrication of both the underpinning quantum hardware~\cite{WangNatphoton2019} and the corresponding electronics for classical readout and control. Together, this enables miniaturisation and mass-manufacture of small quantum devices\textemdash such as quantum communication nodes~\cite{GisThewNatPhoton07}, quantum sensors~\cite{PirandolaNPhoton2018} and sources of randomness~\cite{HerroroRMP2017}\textemdash and promises the precision and scale of fabrication required to assemble useful quantum computers~\cite{LaddNature2010}. Here we combine CMOS compatible silicon and germanium-on-silicon nano-photonics with silicon-germanium integrated amplification electronics to improve performance of on-chip homodyne detection of quantum light. We observe a 3~dB bandwidth of 1.7~GHz, shot-noise limited performance beyond 9~GHz and minaturise the required footprint to 0.84~$\textrm{mm}^2$. We use the device to observe quantum squeezed light, from 100~MHz to 9~GHz, generated in a lithium niobate waveguide. This demonstrates that an all-integrated approach yields faster homodyne detectors for quantum technology than has been achieved to-date and opens the way to full-stack integration of photonic quantum devices.
}

Silicon photonics is emerging as a sophisticated chip-platform to develop quantum technology.
It is being used to realise general purpose programmable networks to process quantum states generated on-chip, with hundreds of individual components~\cite{WangNatphoton2019}, as well as more specific tasks such as 
quantum random number generation~\cite{raffaelli2018homodyne} and chip-to-chip quantum communications~\cite{zh-natphot-Adv-Online}.
However, silicon quantum photonics is yet to exploit integration with monolithic complementary metal-oxide-semiconductor (CMOS) electronics --- classical interface hardware for integrated quantum photonics is still being implemented with large-footprint discrete electronics that limit device scalability and performance. Fully exploiting CMOS compatibility is of particular importance for reducing total device footprint and where high performance classical control and readout resources must scale with the underpinning quantum hardware.

Homodyne detectors are hardware that are routinely used for sensitive measurement of light. Applications include coherent Ising machines~\cite{McMahon614}, dual-comb spectroscopy~\cite{millot2016frequency}, quantum-secured communication~\cite{zh-natphot-Adv-Online},
quantum state and process tomography~\cite{RevModPhys.81.299}, quantum computing~\cite{PhysRevLett.97.110501}, random number generation~\cite{ga-natphot-4-711}, quantum teleportation~\cite{Furusawa706} and ultra-sensitive interferometry, such as gravitational wave astronomy~\cite{PhysRevLett.123.231107}. 
Homodyne detectors are also needed for recently proposed optical neural networks performing below the Landauer limit~\cite{PhysRevX.9.021032}. But the speed performance of many applications is limited by the use of non-monolithic photonic and electronic circuits. In particular, detector bandwidth can define the clock-rate of continuous variables (CV) quantum computers~\cite{Larsen369,Asavanant373}, the secure-key exchange rate for CV quantum communications~\cite{zh-natphot-Adv-Online}, and the rate of detection and heralded preparation of pulsed quantum states~\cite{zavatta2004tomographic}.

Integrated photodiodes promise to increase homodyne detector bandwidth ---
tens of GHz speed performance is already being exploited for classical silicon photonic interconnects~\cite{Virot:GEPD}. However, for shot-noise limited applications --- that includes quantum technology --- the integrated approach has so far been limited to 150~MHz~\cite{raffaelli2018homodyne,zh-natphot-Adv-Online}, because the transimpedance amplifiers (TIA)---needed to amplify the weak photocurrent subtraction of the two photodiodes with low noise---were implemented with discrete electronic components. This introduces parasitic capacitance that increases noise of the circuit and limits the bandwidth of the overall detector, thus negating the $\sim$10s~GHz raw-performance of integrated photodiodes.

\begin{figure*}[htp!]
    \centering
    \includegraphics[width=1\linewidth]{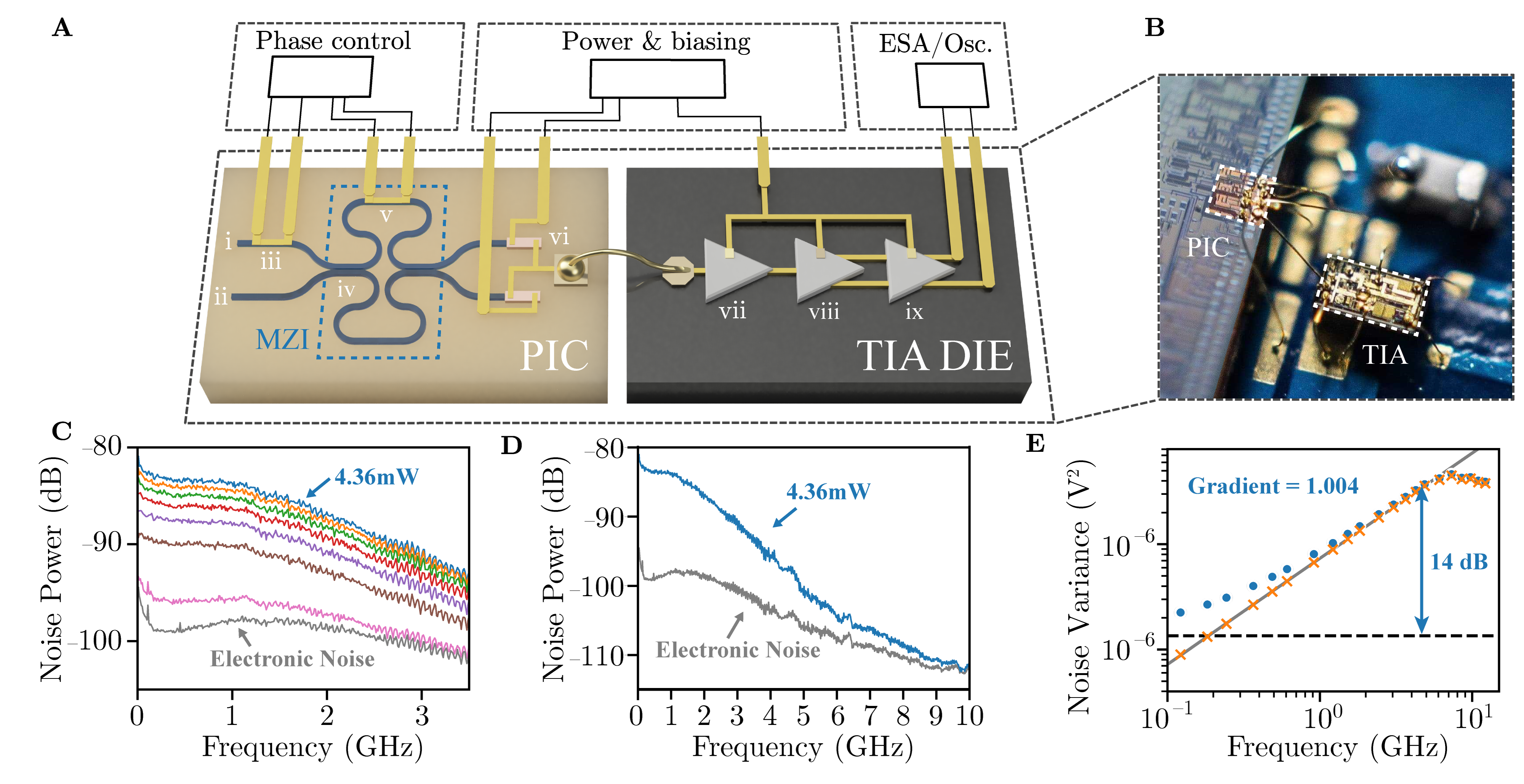}
    \caption{Device schematic and characterisation. \textbf{A} Illustration of the PIC and MAX3277 TIA. The homodyne signal is recorded with a Keysight N9020B electronic spectrum analyser (ESA) and Keysight DSOS604A oscilloscope (Osc.). \textbf{B} Photograph of the homodyne detector device mounted onto a PCB. \textbf{C} Spectral response of the detector with increasing LO powers from bottom to top in the range \{0 mW, 0.12 mW, 0.85 mW, 1.58 mW, 2.30 mW, 3.00 mW, 3.69 mW, 4.36 mW\}, corrected for PIC insertion loss. \textbf{D} spectral response of detector across 10 GHz of spectrum analyser bandwidth. \textbf{E} Raw (dots) and electronic-noise-subtracted (crosses) signal variance with different CW LO powers integrated over a bandwidth of 1.7 GHz. The black line represents the noise floor of the detector. The grey line is fitted over all but the last five data points where the photodiodes begin to saturate.}
    \label{fig:detcharacterisation}
\end{figure*}

Here we show that combining microelectronics with silicon nanophotonics enables enhanced performance homodyne detectors for measuring non-classical light whilst miniaturising the detector's footprint to sub-$\textrm{mm}^2$, all in CMOS-compatible technology for scale.
Our homodyne detector is illustrated in Fig.~1 and implements, in a photonic integrated circuit (PIC) chip architecture, all the integrated linear optics~\cite{ma-natphot-316-9} together with on-chip photodiodes~\cite{raffaelli2018homodyne} needed for a standard homodyne detector.
The silicon PIC is designed in-house and fabricated by IMEC Foundry Services using their ISiPP50G process \cite{SOIplatformsreview}. A local oscillator (LO)---used to extract phase information from a signal field---is guided into 450~$\times$~220nm single mode waveguides via a vertical grating coupler before input (ii). The signal measured is input into (i). The phase between signal and LO is controlled with a thermal phase shifter (iii) and the two fields interfere at a beamsplitter implemented with a variable Mach-Zehnder interferometer (MZI) comprising two directional couplers (iv) and a second thermal phase shifter (v) to precisely tune reflectivity.
The chip exhibits crosstalk between the LO and MZI phase shifters of approximately 0.9\% which is mitigated by locking the MZI to 50\% reflectivity using a PID feedback routine (see Supplementary Information section 7 for details). Two waveguide-integrated germanium photodiodes (vi)\textemdash approximately $100~\mu$m in length with foundry target responsivities of 1.1~A/W and 50~$\Omega$ bandwidths of 17~GHz\textemdash are coupled to the two outputs of the MZI. The two photodiodes are connected together in a balanced configuration on the metallisation layer of the PIC with output ports accessed by bond pads. 
The PIC is wire bonded directly to a commercial TIA die: an unpackaged MAX3277 (Maxim Integrated) with typical transimpedance gain of 
$3.3$~k$\Omega$ and fabricated in a Si-Ge bipolar process~\cite{MaximIntegrated2005MaximDatasheet}. The die consists of an initial TIA stage (vii), followed by a voltage amplifier (viii) and output buffer amplifier (ix).
Direct interfacing is used to access the speed performance of the integrated photodiodes and the TIA by minimising the parasitic capacitance and inductance at the input to the amplifier~\cite{MaximIntegrated2005MaximDatasheet}. 
Both PIC and TIA die are mounted on a custom-designed printed circuit board (PCB) for biasing (Fig.~1~B, Supplementary Information section 1 for details).

\begin{figure*}[tp!]
    \centering
    \includegraphics[width=1\linewidth]{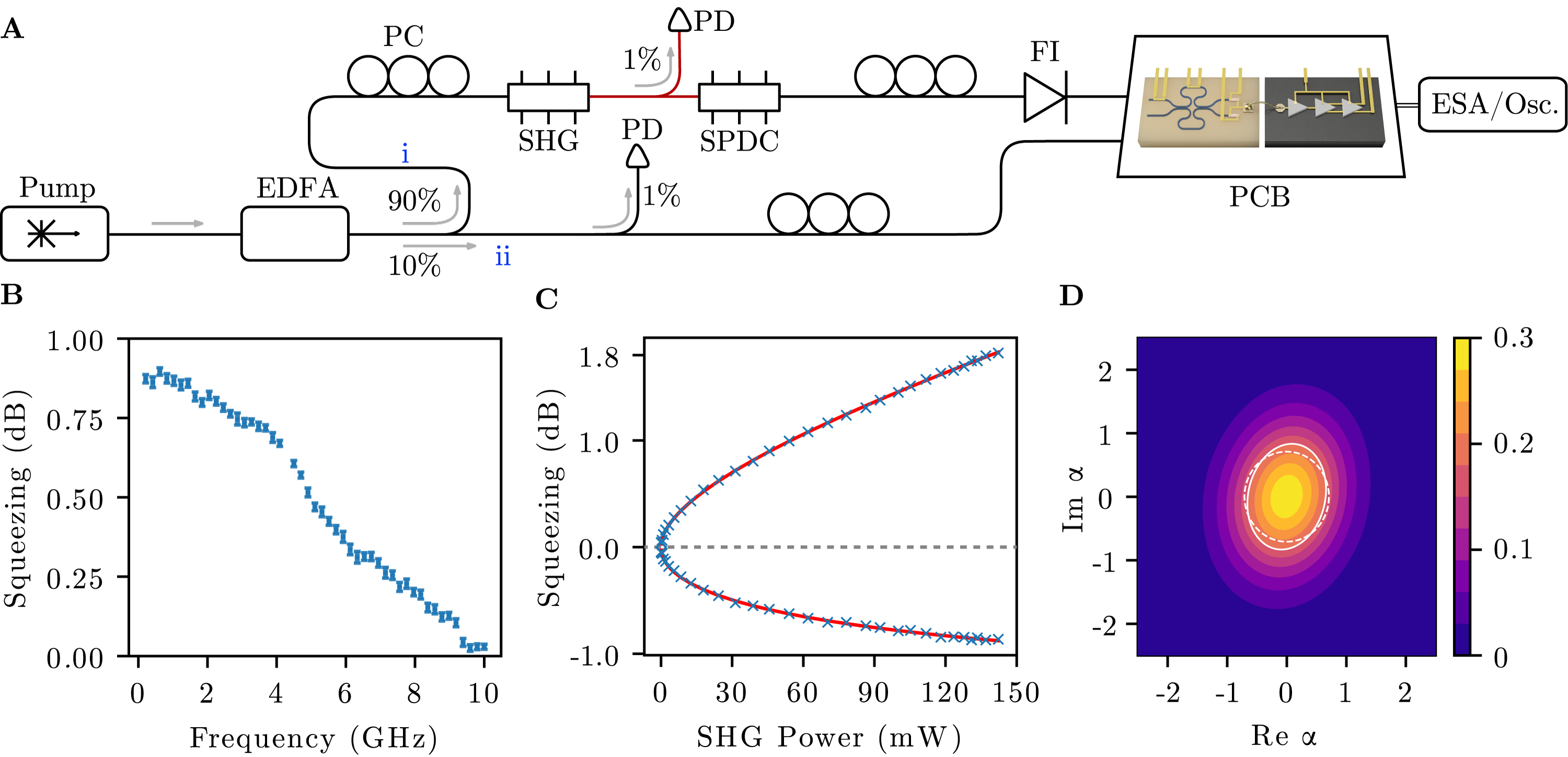}
    \caption{Measurement of a fibre coupled squeezing source using the integrated detector. \textbf{A} Experimental setup. A tunable CW pump laser centred at 1560.61~nm is amplified with an erbium doped fibre amplifier (EDFA) before splitting into squeezing (i) and local oscillator (ii) paths. The polarisation of the light in both paths is set (PC) to optimise SHG and squeezed light generation via SPDC, and PIC input coupling. Both SHG and SPDC are implemented with a lithium niobate periodically-poled ridge waveguide (NTT Electronics WH-0780-000-F-B-C-M). Attenuation of the 90\% path controls the SHG pump power independent of the LO. Power is monitored (Thorlabs S121C, S154C. PD in figure) in each arm with 99:1 fibre splitters. A fibre isolator (FI) acting as a filter prevents residual 780nm light from reaching the homodyne detector. Noise powers are collected over 1~s with an ESA in zero-span mode set to 8~MHz resolution bandwidth. \textbf{B} Estimated squeezing levels across detector bandwidth (see Supplementary Information section 7 for method of estimation). One data point has been removed at 4.29 GHz due to environmental RF interference. The drop in measured squeezing at higher frequencies is expected to be due to the reduction of shot-noise clearance which adds an effective loss (Supplementary Information 8). \textbf{C} Measured squeezing and anti-squeezing levels (blue crosses) with SHG power monitored using a 99:1 beamsplitter between PPLN modules. Eqn~(\ref{eqn:powerfit}) has been fit to find total detection efficiency $\eta_{total}$ and $\mu$ (red line). Error bars are smaller than data points. \textbf{D} Reconstructed Wigner function of the squeezed vacuum integrated from 0-2~GHz. Solid and dashed white lines represent $1/e$ points of the squeezed and vacuum states respectively.}
    \label{fig:sqexptdiag}
\end{figure*}

The utility of a homodyne detector for quantum applications can be measured by its frequency response (including its ability to resolve the shot-noise of a bright local oscillator field above the dark noise of the detector), common-mode rejection ratio (CMRR), and overall quantum efficiency. Characterisation of the detector is performed with no light (vacuum) coupled to input (i). The frequency response and shot-noise clearance as a function of frequency determines the time mode of quantum states that may be accurately characterised by the detector---for our detector, this is shown in Figs.~1~C~and~D. A common bandwidth metric is the 3-dB cutoff frequency of the shot-noise, however measurements can be made beyond this provided there is sufficient shot-noise clearance. Accounting for frequency dependent losses in the coaxial radio frequency (RF) connecting cables, we obtain a 3-dB bandwidth of 1.7~GHz by fitting measured data with a 
Butterworth response (Supplementary Information section 2). Fig.~1~D shows that the detector enables shot-noise limited measurements to beyond 9~GHz.

The CMRR determines how well the photocurrent subtraction suppresses common-mode signals, such as classical laser amplitude noise. CMRR is set at DC by the accuracy of the beamsplitter reflectivity and matching of photodiode responsivities but is degraded at higher frequencies by optical and electronic path length imbalances~\cite{Painchaud:09}. The inherent mode-matching of integrated optics, as well as the precise path-length matching afforded to integrated circuits, enables a high CMRR to be achieved through the tuning of a single thermal phase-shifter. By amplitude modulating the LO and varying the MZI between balanced and unbalanced configurations, we measure a CMRR of $>52$~dB at 1~GHz (we attribute this limit to the dynamic range of the detector\textemdash see Supplementary Information 3). 

Quantum efficiency is determined by losses in the detector's optics, the efficiency of the photodiodes, and its shot-noise clearance. We cannot probe the individual photodiode efficiencies without the presence of coupled waveguides, so instead measure the combination of optical loss and efficiency to be $\eta_{det}$ $\sim$88\% (Supplementary Information 4). To measure the shot-noise clearance of our detector and verify that the photocurrent noise increases in power linearly with the LO power\textemdash a signature that the detector is shot-noise-limited\textemdash we plot the variance of the detector output as a function of input power on a bi-logarithmic scale (Fig.~1~E). In the linear, unsaturated regime, we obtain a maximum shot-noise clearance of 14~dB and noise-subtracted gradient of $1.004 \pm 0.003$. This gives an efficiency contribution of 96\% and therefore a total quantum efficiency of the detector of $0.88\times 0.96 = 84\%$ for measuring states guided in the PIC. 

Many quantum optics applications involve measuring squeezed light and states derived from squeezing by non-Gaussian operations~\cite{RevModPhys.81.299}.
We apply our device to measure broadband squeezed light generated from a fibre-coupled periodically-poled lithium niobate (LN) ridge waveguide chip (Fig.~2~A). The source~\cite{Kaiser:16} is composed of an amplified telecom pump laser, a frequency doubling stage (SHG) and a parametric downconverter (SPDC) --- we launch the generated squeezed vacuum 
into the second input of the chip provided for the signal input. Fig.~2~B shows the measured squeezing across the full detector bandwidth, directly observing an average of $0.87~\pm~0.01$~dB of squeezing over the 3-dB bandwidth, diminishing to electronic noise at 9.2~GHz. To measure the squeezing and anti-squeezing shown in Fig.~2~C, the local oscillator phase is scanned using the on-chip phase shifter (Supplementary Information 5) at 5~Hz. This is repeated for a range of SHG powers (directly monitored between modules). We fit the data with quadrature variances described by
\begin{equation}
    \label{eqn:powerfit}
    \Delta X^2_{\frac{\text{MAX}}{\text{MIN}}} = \eta_{total} \exp{(\pm2r)} + (1 - \eta_{total}),
\end{equation}
with overall setup efficiency $\eta_{total}$ and $r\!=\!\mu\sqrt{P_{SHG}}$~\cite{Kaiser:16}, for anti-squeezing (MAX) and squeezing (MIN). From this we obtain $\hat{\eta}_{total} = 0.28 \pm 0.01$ and $\mu = 0.044 \pm 0.002$ mW\textsuperscript{-1/2}. 

To account for the overall squeezing measurement efficiency, $\eta_{total}$, we identify all the sources of loss. From Ref.~\cite{Kaiser:16} we assume $\sim 0.83$ transmission through the ridge waveguide and $\sim 0.8$ coupling between the SPDC module and output fibre, followed by $0.85$ transmission measured through the fibre isolator, polarisation controller and fibre couplers. The coupling into the photonic chip is measured to be $-2.1$~dB ($=$0.62) per grating coupler and is monitored throughout the experiment. The measured photodiode quantum efficiency of $\eta_{det}=88\%$ and shot-noise contribution of $\sim 96\%$ means the device acts as a fibre-coupled module with detection efficiency of $0.51$. We estimate the overall detection efficiency of the squeezing experiment as $\eta_{total} \approx 0.292$, which is in close agreement to $\hat{\eta}_{total} = 0.28 \pm 0.01$. After correction for $\eta_{total}$, we estimate a squeezing value of $3.26 \pm 0.15$ dB at the SPDC module output. 

To demonstrate the device's capability to reconstruct input quantum states with phase dependence, we perform homodyne tomography, on-chip, of the detected squeezed vacuum. Using the input thermal phase shifter (iii, Fig.~1), we scan the relative phase between LO and signal by applying a 100~Hz periodic voltage\textemdash faster than the slow thermal phase fluctuations arising in the fibres due to the environment\textemdash and acquire $2 \times 10^7$ phase-quadrature pairs over an integrated bandwidth of 2~GHz (see Supplementary Information 9 for more information). From these, we obtain a reconstruction of the quantum state by applying an iterative maximum likelihood estimation algorithm in a five photon truncated Fock space~\cite{lvovsky2004iterative}. The Wigner function of the reconstructed density matrix is shown in Fig.~2~D.

We have demonstrated that silicon-photonic homodyne detector performance is increased by interfacing with a micro-electronics amplifier. The achieved 3~dB bandwidth of 1.7~GHz is an order of magnitude greater than the previous fastest integrated silicon-germanium demonstration~\cite{raffaelli2018homodyne}, while our measurements of shot-noise above 9~GHz indicates potential for an order of magnitude wider bandwidth than the state of the art in standard homodyne detection~\cite{12Ghzhomodyne}. 
For future use as fibre coupled modules, higher efficiency is possible through improved coupling to optical fibre---Ref.~\cite{ding2014gc} exceeds 85\% coupling efficiency. Our measurement of squeezed light from LN waveguide at sideband frequencies up to 9~GHz suggests benefit from hybrid integration with LN~\cite{Lenzinieaat9331} and silicon nitride~\cite{PhysRevApplied.3.044005}
platforms being developed for integrated CV quantum optics, while the sub-mm$^2$ footprint enables the fabrication of arrays of homodyne detectors for measuring weak fields in multiple spatial modes. We suggest integration with germanium-silicon single photon detectors~\cite{vi-natcomm-10-1086} could provide all the detection methods needed on one silicon platform for non-Gaussian quantum optics~\cite{RevModPhys.81.299}.

\textbf{Data availability} The data that support the plots within this paper and other findings of this study are available from the corresponding author upon
reasonable request.


\textbf{Acknowledgements} The authors are grateful to A Crimp, M Loutit and G Marshall for technical assistance and D Mahler for helpful discussion. This work was supported by EPSRC programme grant EP/L024020/1, EPSRC UK Quantum Technology Hub QUANTIC (EP/M01326X/1), EPSRC Quantum Technology Capital fund: QuPIC (EP/N015126/1) and the Centre for Nanoscience and Quantum Information (NSQI). JF acknowledges support from EPSRC Quantum Engineering Centre for Doctoral Training EP/LO15730/1 and Thales Group. EJA acknowledges support from EPSRC doctoral prize (EP/R513179/1). ST, VDA, and FB acknowledge financial support from the European Union by means of the Fond Europ\'{e}en de d\'{e}veloppement regional (FEDER) through the project OPTIMAL, the Agence Nationale de la Recherche (ANR) through the projects HyLight (ANR-17- CE30-0006-01) and SPOCQ (ANR-14-CE32-0019), and the French government through the program ``Investments for the Future" under the Universit\'{e} C\^ote d`Azur UCA-JEDI project (Quantum@UCA) managed by the ANR (ANR-15-IDEX-01). JCFM acknowledges support from an EPSRC Quantum Technology Fellowship (EP/M024385/1) and an ERC starting grant ERC-2018-STG 803665.


\end{document}